\documentclass[%
 reprint,
 superscriptaddress,
 amsmath,
 amssymb,
 aps,
 pra,
 floats,
 floatfix
]{revtex4-2}

\usepackage{graphicx} 
\usepackage[breaklinks=true]{hyperref} 
\usepackage{physics}
\usepackage{bbold}
\usepackage{mathtools}
\usepackage{siunitx}
\usepackage[caption=false]{subfig}

\DeclareSIUnit{\au}{{a.u.}}

\makeatletter
\let\save@mathaccent\mathaccent
\newcommand*\if@single[3]{%
  \setbox0\hbox{${\mathaccent"0362{#1}}^H$}%
  \setbox2\hbox{${\mathaccent"0362{\kern0pt#1}}^H$}%
  \ifdim\ht0=\ht2 #3\else #2\fi
  }
\newcommand*\rel@kern[1]{\kern#1\dimexpr\macc@kerna}
\newcommand*\widebar[1]{\@ifnextchar^{{\wide@bar{#1}{0}}}{\wide@bar{#1}{1}}}
\newcommand*\wide@bar[2]{\if@single{#1}{\wide@bar@{#1}{#2}{1}}{\wide@bar@{#1}{#2}{2}}}
\newcommand*\wide@bar@[3]{%
  \begingroup
  \def\mathaccent##1##2{%
    \let\mathaccent\save@mathaccent
    \if#32 \let\macc@nucleus\first@char \fi
    \setbox\z@\hbox{$\macc@style{\macc@nucleus}_{}$}%
    \setbox\tw@\hbox{$\macc@style{\macc@nucleus}{}_{}$}%
    \dimen@\wd\tw@
    \advance\dimen@-\wd\z@
    \divide\dimen@ 3
    \@tempdima\wd\tw@
    \advance\@tempdima-\scriptspace
    \divide\@tempdima 10
    \advance\dimen@-\@tempdima
    \ifdim\dimen@>\z@ \dimen@0pt\fi
    \rel@kern{0.6}\kern-\dimen@
    \if#31
      \overline{\rel@kern{-0.6}\kern\dimen@\macc@nucleus\rel@kern{0.4}\kern\dimen@}%
      \advance\dimen@0.4\dimexpr\macc@kerna
      \let\final@kern#2%
      \ifdim\dimen@<\z@ \let\final@kern1\fi
      \if\final@kern1 \kern-\dimen@\fi
    \else
      \overline{\rel@kern{-0.6}\kern\dimen@#1}%
    \fi
  }%
  \macc@depth\@ne
  \let\math@bgroup\@empty \let\math@egroup\macc@set@skewchar
  \mathsurround\z@ \frozen@everymath{\mathgroup\macc@group\relax}%
  \macc@set@skewchar\relax
  \let\mathaccentV\macc@nested@a
  \if#31
    \macc@nested@a\relax111{#1}%
  \else
    \def\gobble@till@marker##1\endmarker{}%
    \futurelet\first@char\gobble@till@marker#1\endmarker
    \ifcat\noexpand\first@char A\else
      \def\first@char{}%
    \fi
    \macc@nested@a\relax111{\first@char}%
  \fi
  \endgroup
}

\usepackage{comment}
\newcommand{\CC}{\text{CC}}
\newcommand{\HF}{\text{HF}}
\newcommand{\tbar}{\widebar{t}}
\newcommand{\Abar}{\widebar{A}}
\newcommand{\Hbar}{\widebar{H}}

\newcommand\T{\rule{0pt}{2.6ex}}       
\newcommand\BS{\rule[-1.3ex]{0pt}{0pt}} 
\newcommand\BM{\rule[-1.8ex]{0pt}{0pt}} 

\begin{document}

\title{Time-dependent coupled cluster theory for ultrafast transient absorption spectroscopy}

\author{Andreas S. \surname{Skeidsvoll}}
\thanks{%
These authors contributed equally to this work.
}%
\affiliation{%
Department of Chemistry, Norwegian University of Science and Technology, 7491 Trondheim, Norway
}%
\author{Alice \surname{Balbi}}
\thanks{%
These authors contributed equally to this work.
}%
\affiliation{%
Scuola Normale Superiore, Piazza  dei  Cavalieri, 7, I-56126, Pisa, Italy
}%

\author{Henrik \surname{Koch}}
\email{Electronic mail: henrik.koch@sns.it}
\affiliation{%
Department of Chemistry, Norwegian University of Science and Technology, 7491 Trondheim, Norway
}
\affiliation{%
Scuola Normale Superiore, Piazza  dei  Cavalieri, 7, I-56126, Pisa, Italy
}

\date{\today}

\begin{abstract}
We present a spin-adapted time-dependent coupled cluster singles and doubles model for the molecular response to a sequence of ultrashort laser pulses. The implementation is used to calculate the electronic response to a valence-exciting pump pulse, and a subsequent core-exciting probe pulse. We assess the accuracy of the integration procedures used in solving the dynamic coupled cluster equations, in order to find a compromise between computational cost and accuracy. The transient absorption spectrum of lithium fluoride is calculated for various delays of the probe pulse with respect to the pump pulse. We observe that the transient probe absorption oscillates with the pump-probe delay, an effect that is attributed to the interference of states in the pump-induced superposition.
\end{abstract}

\maketitle

\section{Introduction}
Recent advances in the field of ultrafast pulse shaping have enabled the generation of broadband few- to sub-femtosecond laser pulses in the near infrared to vacuum ultraviolet spectral ranges \cite{Hassan2016,Galli:19,Fabris2015}. These ultrashort pulses open the possibility to study valence electron dynamics of molecules, on time scales shorter than times characteristic for nuclear dynamics. Also, the generation of intense isolated soft X-ray free electron laser pulses with sub-femtosecond temporal widths has recently been achieved \cite{Duris2020}. This paves the way for attosecond-resolved core-level spectroscopy at high intensities and repetition rates.

Core excitations are typically local to specific atoms, and are sensitive to their electronic environment \cite{Wolf2017}. The associated attosecond-resolved transient absorption can thus be used to observe superpositions of valence-excited states from the point of view of a specific atomic site, provided that the superposition is of a certain degree of coherence \cite{Goulielmakis2010}. In the short-pulse limit, the energy-integrated absorption of a core-exciting pulse is indicative of the electronic hole density in the valence region around the nucleus of the specific atom \cite{Dutoi2013,Dutoi2014}. For sub-femtosecond pulses outside this limit, the relationship between the pump-induced charge migration and the resultant transient absorption of the probe pulse is more complex. Thus more complete theoretical models are necessary for guiding the pump-probe experiments and for interpreting ensuing results.

Provided that the transient absorption of a probe pulse can be modelled and understood, the valence-level pump and subsequent core-level probe by ultrafast pulses can then be used to investigate the valence electron response of molecules \cite{Dutoi2013,Dutoi2014}. A refined conceptual understanding of this response will shed light on processes occurring in nature, such as photosynthesis and eyesight, and be used for the advancement of technological applications, such as photovoltaics and photocatalysis.

Non-perturbative modelling of electron dynamics for ultrafast laser-matter interactions offers certain advantages: the models are applicable for a large range of field intensities \cite{Nascimento2016}, and the interaction between a molecule and ultrashort pulses resembles experimental setups in a more natural way.

Electron correlation is often important for a qualitative and quantitative description of many-electron systems. The full configuration interaction (FCI) model is computationally impracticable in most situations \cite{OLSEN1990463}, and thus we advocate the use of coupled cluster theory in this paper. Other methods have been used to describe electron dynamics, such as real-time density functional theory (DFT) \cite{doi:10.1063/1.1479349,doi:10.1063/1.4953039}. However, DFT methods are limited by the accuracy of the exchange correlation functionals, and thus could lead to misinterpretations. Several implementations of real-time coupled cluster models have been developed in the past, including approaches based on the time-dependent coupled cluster (TDCC) equations derived by Koch and J{\o}rgensen \cite{doi:10.1063/1.458814,doi:10.1063/1.4718427,Pigg2012,doi:10.1063/1.5142276}, and approaches based on equation of motion (EOM) theory \cite{Sonk2011,Sonk2011a,doi:10.1080/00268976.2012.675448,Nascimento2016,Nascimento2017,Koulias2019,doi:10.1063/1.5125494,doi:10.1063/1.5117841}. These models offer an accurate description of dynamic correlation, and static correlation in excited states. Needless to say, the coupled cluster models are also inherently size extensive and intensive \cite{doi:10.1063/1.457710}. This while keeping the polynomial scaling of the computational costs with respect to system size. 

A spin-unrestricted time-dependent coupled cluster singles and doubles (TDCCSD) model was recently implemented by Pedersen and Kvaal, and used to calculate the absorption spectra of helium and beryllium irradiated by ultrashort pulses at various intensities \cite{doi:10.1063/1.5085390}. Even above the perturbative limit, the TDCCSD spectra show promising correspondence with spectra calculated with time-dependent FCI. The authors also noted that the Lagrangian time-dependent equations have a Hamiltonian structure, well suited for the use of symplectic integrators.

In this work, we will continue the discussion of TDCC models, by presenting a spin-adapted TDCC model of ultrafast transient absorption spectroscopy. Applied to closed-shell molecules interacting with laser pulses within the dipole approximation, this model offers equivalent results as its spin-unrestricted counterparts, with lower computational costs. The reduced cost implies that larger molecules can be studied within this model, making progress towards the accurate modelling of correlated dynamics in interesting photoactive molecules.

This paper is organized as follows. In section \ref{sec:theory} we present the theory underlying the TDCC model and discuss a generalization of Ehrenfest's theorem in this framework. We also describe how absorption spectra are calculated. In section \ref{sec:results}, we optimize the different parameters used in TDCC calculations, and illustrate this for the LiH molecule. The model is applied to transient absorption of the LiF molecule. Final remarks are given in section \ref{sec:conclusion}.
%
\section{Theory} \label{sec:theory}
%
\subsection{Spin-adapted coupled cluster method}
%
An accurate account of the electron correlation in molecules is offered by coupled cluster models, in which the time-independent wave function can be written as
\begin{equation}
    \label{eq:ket}
    \ket{\CC} = e^{T} \ket{\HF},
\end{equation}
where $\ket{\HF}$ is the closed-shell Hartree-Fock reference determinant and $T$ is the spin-adapted cluster operator. The cluster operator is defined as a linear combination of singlet excitation operators $\tau_\mu$,
\begin{equation}
    T = \sum_{\mu>0} t_{\mu} \tau_{\mu}.
\end{equation}
The expansion coefficients $t_{\mu}$ are referred to as the amplitudes. The operator $T$ is usually truncated at a given level of excitation, for instance after single excitations gives the coupled cluster singles (CCS) model, after double excitations gives the coupled cluster singles and doubles model (CCSD), and so on.

In the Lagrangian formulation of coupled cluster theory, which satisfies the Hellman-Feynman theorem, the dual state corresponding to the $\ket{\CC}$ state is \cite{doi:10.1063/1.457710}
\begin{equation}
    \bra{\Lambda} = \bigg(\!\bra{\HF} + \sum_{\nu>0} \tbar_{\nu} \bra{\nu}\!\bigg)e^{-T},
\end{equation}
where the linear expansion coefficients $\tbar_\nu$ will be referred to as the (Lagrange) multipliers. The level of excitations is truncated at the same level as the excitations in the cluster operator. We note that the $\ket{\CC}$ state and its dual state $\bra{\Lambda}$ are biorthonormal, $\bra{\Lambda}\ket{\CC} = 1$.

In this formulation, the expectation values of operators are given as
%
%
\begin{equation}
    \begin{split}
        \ev{A} &= \mel{\Lambda}{A}{\CC} \\
        &= \bigg(\!\bra{\HF} + \sum_{\nu>0} \tbar_{\nu} \bra{\nu}\!\bigg)\Abar\ket{\HF}
    \end{split}
\end{equation}
where the similarity transformed operator is defined as
\begin{equation}
    \Abar = e^{-T}Ae^{T}.
\end{equation}
The amplitudes and multipliers that parameterize the ground state are determined from \cite{doi:10.1002/9781119019572.ch13}
\begin{gather}
    \label{eq:groundstate}
    \mel{\mu}{\Hbar}{\HF} = 0, \\
    \mel{\Lambda}{\comm{H}{\tau_{\mu}}}{\CC} = 0,
\end{gather}
and the corresponding ground state energy $E_{CC}$ is given by
\begin{equation}
    \begin{split}
        E_{CC} &= \mel{\Lambda}{H}{\CC} \\
        &= \mel{\HF}{H}{\CC},
    \end{split}
\end{equation}
where we have used Eq. \eqref{eq:groundstate} to eliminate the multiplier contribution.

\subsection{Time-dependent coupled cluster methods}
In order to allow for time-dependence in the description, the coupled cluster state is parameterized as \cite{doi:10.1063/1.458814}
\begin{equation}
    \ket{\CC(t)} = e^{T(t)}\ket{\HF}e^{i\epsilon(t)},
\end{equation}
and the corresponding dual state as
\begin{equation}
    \bra{\Lambda(t)} = \bigg(\!\bra{\HF} + \sum_{\nu>0} \tbar_{\nu} (t) \bra{\nu}\!\bigg) e^{-T(t)}e^{-i\epsilon(t)}.
\end{equation}
The amplitudes $t_{\mu}$ and multipliers $\tbar_{\mu}$ now explicitly depend on time, while the excitation operators $\tau_{\mu}$ are still time-independent. An overall time-dependent phase $\epsilon(t)$ has also been introduced.

The equation describing the time evolution of the amplitudes $t_{\mu}(t)$ is obtained from the time-dependent Schr\"{o}dinger equation for the $\ket{\CC}$ state, by projecting onto the corresponding excited determinant $\bra{\mu}$. This gives the differential equation
\begin{equation}
    \label{eq:ddtamplitudes}
    \dv{t_{\mu}(t)}{t} = -i\mel{\mu}{\Hbar(t)}{\HF}.
\end{equation}
The equation describing the time evolution of the multipliers $\tbar_{\mu}(t)$ is obtained by projecting the time-dependent Schr\"{o}dinger equation for the dual state $\bra{\Lambda (t)}$ onto the excited determinants $\ket{\nu}$, giving the differential equation
\begin{equation}
    \label{eq:ddtmultipliers}
    \dv{\tbar_{\nu}(t)}{t} = i\Big(\!\bra{\HF}+\sum_{\mu>0}\tbar_{\mu}(t)\bra{\mu}\!\Big)\comm*{\Hbar(t)}{\tau_{\nu}}\!\ket{\HF}.
\end{equation}
The equation for the phase $\epsilon (t)$ is determined by projection onto the $\ket{\HF}$ state
\begin{equation}
    \label{eq:ddtphase}
    \dv{\epsilon(t)}{t} = -\mel{\HF}{\Hbar(t)}{\HF}.
\end{equation}
Detailed derivations can be found in reference \cite{doi:10.1063/1.458814}. In this framework, the time-dependent expectation value of a generic operator $A(t)$ is defined as
\begin{equation}
    \label{eq:tdexpval}
    \begin{split}
        \ev{A(t)} &= \mel{\Lambda (t)}{A (t)}{\CC(t)},
    \end{split}
\end{equation}
where $\braket{\Lambda(t)}{\CC(t)} = 1$.
\subsection{A generalized Ehrenfest's theorem and conserved quantities in TDCC}
For ease of notation, we suppress the explicit time dependence in this section. Ideally, observables calculated in truncated TDCC should have the same properties as in the untruncated case; in order to give a faithful representation of the physical system. In this context, we derive a generalized Ehrenfest's theorem for truncated TDCC (the detailed derivation is given in Appendix A). We obtain the equation
\begin{equation}
    \label{eq:ehrenfest}
    \begin{split}
        \dv{t}\mel{\Lambda'}{A}{\CC} &= i\mel{\Lambda'}{He^{T'}P_{n}e^{-T'}A}{\CC} \\
        &\phantom{{}={}} -i\mel{\Lambda'}{Ae^{T}P_{n}e^{-T}H}{\CC} \\
        &\phantom{{}={}}+ \mel{\Lambda'}{\pdv{A}{t}}{\CC},
    \end{split}
\end{equation}
where the left $\bra{\Lambda'}$ state and the right $\ket{\CC}$ state are independent solutions to the projected time-dependent Schr\"{o}dinger equation. The projection operator $P_{n}$ of maximum excitation level $n$ is defined as
\begin{equation}
    \label{eq:projection}
    P_{n} = \ket{\HF}\!\bra{\HF} + \sum_{\mu>0}^{n}\ket{\mu}\!\bra{\mu},
\end{equation}
and in untruncated TDCC, $P_{n} = \mathbb{1}$. From Eq. \eqref{eq:ehrenfest} we can see that, in untruncated TDCC,
\begin{equation}
    \label{eq:ehrenfest2}
    \dv{t}\mel{\Lambda'}{A}{\CC} = i\mel{\Lambda'}{\comm{H}{A}}{\CC} + \mel{\Lambda'}{\pdv{A}{t}}{\CC},
\end{equation}
regardless of the initial values of the amplitudes, multipliers and phases.

In truncated TDCC, the projection operator cannot in general be replaced by the identity operator, and hence Eq. \eqref{eq:ehrenfest} cannot be simplified further. Still, some conservation laws from untruncated TDCC apply under certain constraints: we see from \eqref{eq:ehrenfest} that the Hamiltonian matrix element $\mel{\Lambda'}{H}{\CC}$ is conserved for a time-independent Hamiltonian operator as long as $T' = T$, regardless of the initial values of the multipliers and phases. The overlap matrix element $\mel{\Lambda'}{\mathbb{1}}{\CC}$ are also conserved for $T' = T$, since $\exp(T){P}_{n}\exp(-T)\mathbb{1}\!\ket{\CC} = \ket{\CC}$ and $\bra{\Lambda'}_{T'=T}\mathbb{1}\exp(T){P}_{n}\exp(-T) = \bra{\Lambda'}_{T'=T}$. In conclusion, we note the energy and overlap conservation for a time-independent Hamiltonian in untruncated TDCC, and in truncated TDCC for $T' = T$.
\subsection{Interaction with an external electromagnetic field}
In the semiclassical approximation, the electronic Hamiltonian for a molecule interacting with an external electromagnetic field can be written as
\begin{equation}
    \label{eq:hamiltonian}
    H(t) = H_0 + V(t),
\end{equation}
where $H_0$ is the time-independent electronic Hamiltonian and $V(t)$ is the operator describing the interaction with the external field.
We choose to express the interaction in the length gauge and dipole approximation, meaning that the electromagnetic field is represented by an electric field,
\begin{equation}
    \begin{split}
        V(t) &= -\vb{d}\vdot\vb*{\mathcal{E}}(t),
    \end{split}
\end{equation}
where $\vb{d}$ is the electric dipole moment operator. Since this operator is a one-electron operator, it can also be expressed in terms of the molecular orbital (MO) dipole moment integrals $\vb{d}_{pq}$ and one-electron singlet excitation operators $E_{pq}$,
\begin{equation}
    \vb{d} = \sum_{pq} \vb{d}_{pq}E_{pq}.
\end{equation}

Since electric fields are additive, the external electric field $\vb*{\mathcal{E}}(t)$ can be written as a linear combination of individual laser pulses,
\begin{equation}
        \vb*{\mathcal{E}}(t) = \sum_{n}\vb*{\mathcal{E}}_{0,n}\cos(\omega_{0,n}(t-t_{0,n}))f_{n}(t),
\end{equation}
where $\vb*{\mathcal{E}}_{0,n}$ is the peak electric field of pulse $n$ in its polarization direction, $\omega_{0,n}$ the carrier frequency and $t_{0,n}$ the temporal midpoint of the pulse, and $f_{n}(t)$ an envelope function that determines its shape.
A commonly used family of envelopes $f_{n}(t)$, that resemble physical laser intensity profiles, are the Gaussian functions. Since Gaussian functions have infinite support, we choose to set them to zero at a finite number $N$ of root-mean-square (RMS) widths $\sigma_{n}$ outside the central time, i.e.
\begin{equation}
    \label{eq:envelope}
    f_{n}(t) =
    \begin{cases}
        e^{-(t - t_{0,n})^{2}/(2\sigma_{n}^{2})}, & a_{n} \le t \le b_{n}, \\
        0, & \text{otherwise},
    \end{cases}
\end{equation}
where $a_{n} = t_{0,n} - N\sigma_{n}$ and $b_{n} = t_{0,n} + N\sigma_{n}$. In addition to resembling physical intensity profiles, a useful feature of Gaussian envelopes is that they give pulses with Gaussian frequency distributions. Hence, these pulses can offer a good compromise between temporal precision and spectral narrowness. This is useful for producing temporally precise electronic transitions within the molecule, while keeping the probability of ionization low.
\subsection{Frequency-resolved transient absorption}
Following the procedure of \cite{Wu_2016}, the energy absorbed during the interaction with the external electromagnetic field can be given by
\begin{equation}
    \Delta E = \int_{-\infty}^{\infty}\dv{E(t)}{t}\dd{t}.
\end{equation}
The time derivative of the expectation value of the Hamiltonian in Eq. \eqref{eq:hamiltonian} can be found through Eq. \eqref{eq:ehrenfest}.
\begin{equation}
    \begin{split}
        \dv{E(t)}{t} &= \dv{t}\mel{\Lambda(t)}{H(t)}{\CC(t)} \\
        &= \mel{\Lambda(t)}{\pdv{H(t)}{t}}{\CC(t)} \\
        &= -\vb{d}(t)\vdot\pdv{\vb*{\mathcal{E}}(t)}{t}
    \end{split}
\end{equation}
where the TDCC dipole moment expectation value is given by
\begin{equation}
    \vb{d}(t) = \mel{\Lambda(t)}{\vb{d}}{\CC(t)}.
\end{equation}
The energy exchanged between the electromagnetic field and the molecule is thus given by
\begin{equation}
    \label{eq:exchangedenergy}
    \Delta E = -\int_{-\infty}^{\infty}\vb{d}(t)\vdot\pdv{\vb*{\mathcal{E}}(t)}{t}\dd{t}.
\end{equation}

Eq. \eqref{eq:exchangedenergy} can be frequency-resolved by inserting the relations between the components ${d}_{i}(t)$ and $\mathcal{E}_{i}(t)$ and their Fourier transforms, ${\widetilde{d}}_{i}(\omega)$ and $\widetilde{\mathcal{E}}_{i}(\omega)$. We use the following convention
\begin{gather}
    f(t) = \frac{1}{\sqrt{2\pi}}\int_{-\infty}^{\infty}\widetilde{f}(\omega)e^{i\omega t}\dd{\omega}, \\
    \widetilde{f}(\omega) = \frac{1}{\sqrt{2\pi}}\int_{-\infty}^{\infty}f(t)e^{-i\omega t}\dd{t}.
\end{gather}
After inserting the relations, the expression
\begin{equation}
    \label{eq:transientenergy}
    \Delta E = \int_{0}^{\infty}\omega S(\omega)\dd{\omega}
\end{equation}
is obtained, where
\begin{equation}
    \label{eq:response}
    S(\omega) = -2 \Im\Big(\widetilde{\vb{d}}(\omega)\vdot\widetilde{\vb*{\mathcal{E}}}^{*}(\omega)\Big) \qc \omega > 0.
\end{equation}
The response function $S(\omega)$ has the opposite sign as in \cite{Wu_2016}, due to different Fourier transform conventions. It represents the absorption per unit frequency at a given frequency, so that positive (negative) $\omega S(\omega)$ equals the amount of energy gained (lost) by the molecule per unit frequency at $\omega$ \cite{Wu_2016}.

The TDCC dipole moment $\vb{d}(t)$ can be found from Eq. $\eqref{eq:tdexpval}$,
\begin{equation}
    \label{eq:dipolemoment}
    \begin{split}
        \vb{d}(t) &=
        \sum_{pq}\mel*{\Lambda(t)}{E_{pq}}{\CC(t)}\vb{d}_{pq} \\
        &= \Big(\!\bra{\HF}+\sum_{\mu>0}\tbar_{\mu}(t)\bra{\mu}\!\Big)\widebar{E}_{pq}(t)\!\ket{\HF}\vb{d}_{pq} \\
        &= \sum_{pq}D_{pq}(t)\vb{d}_{pq},
    \end{split}
\end{equation}
where $D_{pq}(t)$ is an element of the standard coupled cluster one-electron density matrix, which can be calculated given the time-dependent amplitudes and multipliers.
\subsection{Initial value problem}
In order to calculate the time-dependent amplitudes and multipliers for the system represented by the Hamiltonian in Eq. \eqref{eq:hamiltonian}, the system is prepared in the ground state at $t = -T$ (before the interaction). The time-dependent amplitudes and multipliers are then propagated by integration of Eqs. \eqref{eq:ddtamplitudes} and \eqref{eq:ddtmultipliers}, until $t = T$ (after the interaction). This is done using Runge-Kutta methods (a general introduction to these methods is given in Appendix B). Once the time-dependent amplitudes and multipliers are calculated, they can be used to calculate evenly sampled values of the TDCC dipole moment with Eq. \eqref{eq:dipolemoment}.

The main Runge-Kutta method used for integration is the explicit Runge-Kutta (ERK) method known as RK4, and referred to as "the best-known fourth-order four-stage ERK method" in \cite{iserles_2008}. In many cases, this method gives a good compromise between accuracy and the number of evaluations for each time step.

The performance of two methods in the family of $\nu$-stage $2\nu$th-order implicit Runge-Kutta (IRK) methods, known as Gauss-Legendre methods, is also assessed. An interesting property of these methods is that they are symplectic, meaning that they often perform well with regards to preserving the energy expectation value of non-interacting Hamiltonian systems. The application of these methods to TDCC methods was discussed in greater detail in the work by Pedersen and Kvaal \cite{doi:10.1063/1.5085390}. The Gauss-Legendre methods that will be considered here are the two-stage fourth-order Gauss-Legendre method (GL4) and the three-stage sixth-order Gauss-Legendre method (GL6).
\subsection{Discrete Fourier transformation of TDCC dipole moment and electric field}
After the dipole moment and electric field have been calculated in $[-T,T]$, a discrete approximation of ${\widetilde{d}}_{i}(\omega)$ and ${\widetilde{\mathcal{E}}}_{i}(\omega)$ can be found from doing the discrete Fourier transform of the time series.

Assuming that the finite and discrete time series are sampled from infinitely extending analytic dipole moment and electric field functions, the time series can equally be represented as the analytic functions modulated by the rectangular window function,
\begin{equation}
    f_{w_{R}}(t) = f(t)w_{\text{R}}(t)
\end{equation}
sampled in $[-T, T]$, where the rectangular window function,
\begin{equation}
    w_{R}(t) =
    \begin{cases}
        1 \qc \abs{t} \le T, \\
        0 \qc \text{otherwise.}
    \end{cases}
\end{equation}
Since the Fourier transform of a windowed function is equal to the convolution of the Fourier transform of the function with the Fourier transform of the window function \cite{1455106},
\begin{equation}
    \widetilde{f}_{w}(\omega) = \widetilde{f}(\omega) \ast \widetilde{w}(\omega),
\end{equation}
the spectral leakage of the peaks in the finite Fourier spectrum will be related to the Fourier transform of the rectangular window function.
In order to reduce the intensity of side-lobes of peaks in the Fourier spectrum \cite{1455106}, the rectangular window can be replaced with a Hann window, by multiplying the sampled values with the Hann function,
\begin{equation}
    w_{H}(t) = \cos[2](\frac{\pi t}{T}),
\end{equation}
before doing the discrete Fourier transform.
\section{Results and discussion} \label{sec:results}
\subsection{Convergence of LiH pump-probe absorption spectra}
In the following, we investigate the convergence properties of the spin-adapted TDCC model of molecular ultrafast pump-probe absorption. The convergence will be assessed with respect to the individual variation of several parameters: the basis set, the size of the time steps, and the integration method. The TDCC method was implemented in the recently released $e^{T}$ program \cite{doi:10.1063/5.0004713}. This program is used for all reported computations.

The higher level coupled cluster methods scale rapidly with the size of the system, and quickly reach the limits of practicability. Therefore, we have chosen lithium hydride (LiH) for the convergence studies. This serves as an elementary example of a closed-shell molecule with atoms of different core excitation frequencies. The electronic charge can migrate between the two atoms, making it an interesting case for examination by pump-probe spectroscopy.
\begin{table}
    \centering
    \begin{ruledtabular}
    \begin{tabular}{cccc}
        & $\sigma$ [\si{\au}]  & $\omega_{0}$ [\si{\electronvolt}] & $E_{0}$ [\si{\au}]\T\BM \\
        \colrule
        LiH pump & \num{20} & \num{3.55247}
        & \num{0.01}\T \\ 
        LiH probe & \num{10} & \num{57.6527}
        & \num{0.1}\BS \\ 
        \colrule
        LiF pump & \num{20} & \num{6.44801}
        & \num{0.01}\T \\ 
        LiF probe & \num{10} & \num{688.018}
        & \num{0.1} 
    \end{tabular}
    \end{ruledtabular}
    \caption{Gaussian RMS width $\sigma$, central angular frequency $\omega_{0}$ and peak electric field strength $E_{0}$ of the LiH and LiF pump and probe pulses. A Gaussian RMS width of \SI{20}{\au} corresponds to a field strength (intensity) full width at half maximum (FWHM) of \SI{1.139}{\femto\second} (\SI{805.5}{\atto\second}) 
    and \SI{10}{\au} to a FWHM of \SI{569.6}{\atto\second} (\SI{402.8}{\atto\second}). 
    A peak electric field strength of \SI{0.01}{\au} corresponds to a peak intensity of \SI{7.019e12}{\watt\per\square\centi\meter}  
    and \SI{0.1}{\au} to a peak intensity of \SI{7.019e14}{\watt\per\square\centi\meter}.  
    Conversions are done from Hartree atomic units using the 2018 CODATA recommended values \cite{codata} and the peak intensity relation $S_{0} = E_{0}^{2}/Z_{0}$, where $Z_{0}$ is the impedance of free space.}
    \label{tab:pulse_parameters}
\end{table}

The lithium atom is placed at the origin, and the hydrogen atom at \SI{-1.59491318}{\angstrom} along the $z$-axis, corresponding to the experimentally measured equilibrium bond length of LiH \cite{cccbdb}. Gaussian envelopes are used for the pump and probe pulses, which are polarized in the $z$-direction. The electric fields of each pulse are temporally truncated at eight RMS widths $\sigma$ from the central time, and thus nonzero only inside this interval (see Eq. \eqref{eq:envelope}). The central frequency of the pump pulse is tailored to the first LiH valence excitation energy, and the central frequency of the probe pulse to the first LiH K-edge excitation energy. These excitation energies are calculated using EOM-CCSD. The core excitations are obtained within the core-valence separation (CVS) approximation \cite{doi:10.1063/1.4935712}. The parameters of the pulses are shown in Table \ref{tab:pulse_parameters}.

The pump pulse is given a central time of $t = \SI{-40}{\au}$ and the probe a central time of $t = \SI{0}{\au}$. The time-dependent dipole moment and electric field are calculated every \SI{0.1}{\au} in the $[\SI{-5000}{\au}, \SI{5000}{\au}]$ interval. Since the system remains in the ground state until the onset of the truncated pump pulse---with the ground state dipole moment---the interaction with the pulses only needs to be calculated in $[\SI{-200}{\au}, \SI{5000}{\au}]$. Subsequently, the Hann windowed components of the dipole moment and electric field are discrete Fourier transformed, and the transient absorption is calculated using Eq. \eqref{eq:response}.

We use the correlation-consistent basis sets of Dunning et al. (cc-pVXZ, X = D, T) \cite{doi:10.1063/1.456153}, that are suitable for describing valence correlation effects in molecules. In some of the calculations, the basis sets are augmented by diffuse functions (denoted by aug-) and/or functions describing core correlation (denoted by C) \cite{doi:10.1063/1.462569}. From now on, we will use a C in round brackets to indicate that core correlation functions are added to the basis set of the heaviest atom in the molecule.
\begin{figure}
    \centering
    \includegraphics[width=3.375in]{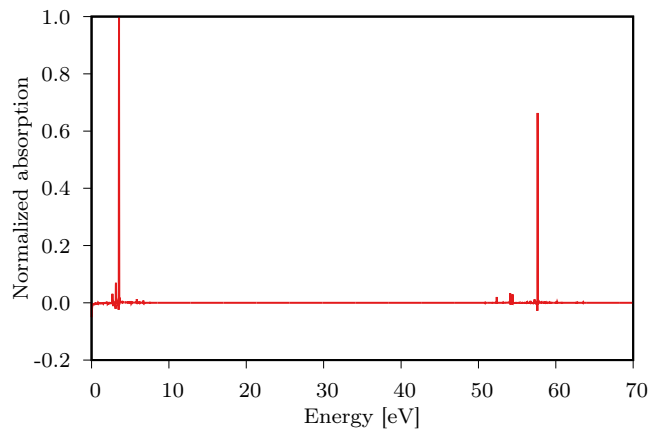}
    \caption{Normalized reference LiH pump-probe absorption, $S'(\omega)$, as a function of energy. The time-dependent dipole moment is calculated using TDCCSD/aug-cc-p(C)VDZ, and integrated with RK4 with \SI{0.005}{\au} time steps.}
    \label{fig:reference}
\end{figure}

The individual variation of the calculation parameters is done with respect to a common reference: TDCCSD/aug-cc-p(C)VDZ, and integrated with RK4 with \SI{0.005}{\au} time steps. The unnormalized reference absorption $S_{\text{ref}}(\omega)$ is used to calculate the normalization factor
\begin{equation}
    \mathcal{N}_{\text{ref}} = \frac{1}{\max_{\omega}\abs{S_{\text{ref}}(\omega)}}.
\end{equation}
This factor is used to normalize all the absorption spectra of the following LiH calculations, by means of
\begin{equation}
    \label{eq:normalization}
    S'(\omega) = \mathcal{N}_{\text{ref}}S(\omega),
\end{equation}
where $S(\omega)$ is calculated with the parameters in question. The normalized deviation of $S'(\omega)$ from a more accurate result $S'_{\text{acc}}(\omega)$ is calculated as
\begin{equation}
    \label{eq:error}
    D'(\omega) = \abs{S'(\omega)-S_{\text{acc}}'(\omega)}.
\end{equation}

The reference absorption spectrum, normalized according to Eq. \eqref{eq:normalization}, is shown in Fig. \ref{fig:reference}. We observe absorption in two energy regions: one corresponding to the valence-exciting pump pulse and the other to the core-exciting probe pulse.

\subsubsection{TDCCS and TDCCSD}
%
\begin{figure}
    \centering
    \includegraphics[width=3.375in]{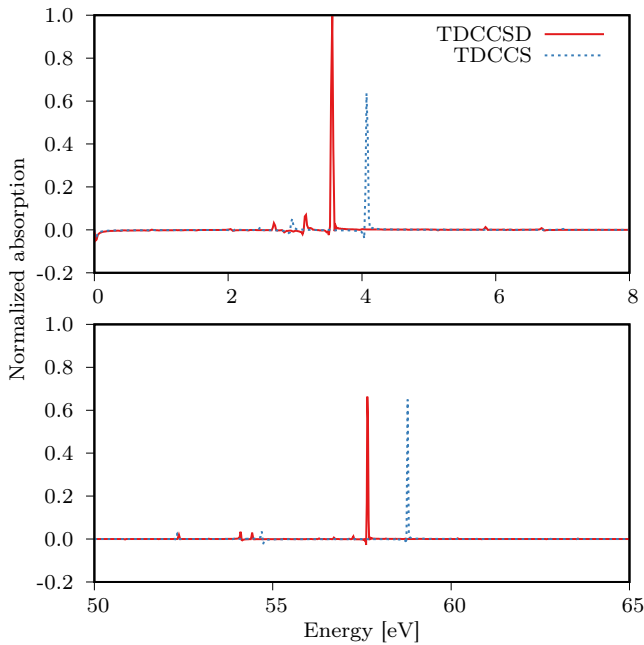}
    \caption{Normalized TDCCSD and TDCCS LiH pump and probe absorption, $S'(\omega)$, as a function of energy. Time-dependent dipole moments are calculated using aug-cc-p(C)VDZ, and integrated with RK4 with \SI{0.005}{\au} time steps.}
    \label{fig:level}
\end{figure}
In Fig. \ref{fig:level}, the normalized reference TDCCSD spectrum is shown together with the normalized time-dependent CCS (TDCCS) spectrum. The two spectra display substantial differences in intensities and positions of the peaks in both the pump and the probe absorption regions. Since TDCCSD includes double and connected quadruple excitations, while TDCCS does not \cite{doi:10.1002/9781119019572.ch13}, this demonstrates that a higher order representation of the correlation is needed to obtain qualitatively correct results for the LiH model system.

\subsubsection{Basis set}
%
\begin{figure}
    \centering
    \includegraphics[width=3.375in]{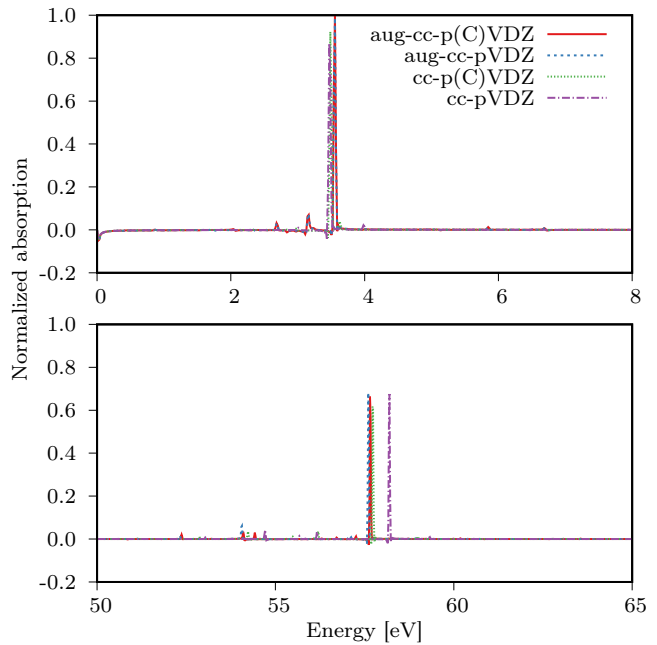}
    \caption{Normalized aug-cc-p(C)VDZ, aug-cc-pVDZ, cc-p(C)VDZ and cc-pVDZ LiH pump and probe absorption, $S'(\omega)$, as a function of energy. Time-dependent dipole moments are calculated using TDCCSD, and integrated with RK4 with \SI{0.005}{\au} time steps.}
    \label{fig:basis}
\end{figure}
In Fig. \ref{fig:basis}, the normalized reference spectrum is shown together with normalized spectra calculated using cc-pVDZ, cc-p(C)VDZ and aug-cc-pVDZ. The inclusion of diffuse functions in the basis sets seems important for representing the dynamics properly. Increasing the basis set from cc-pVDZ to aug-cc-pVDZ shifts the peaks in both the pump and the probe absorption regions. This is consistent with the concept of the pump pulse forcing electrons to the outer valence regions of the molecule, which is better represented with diffuse functions.

Furthermore, comparing cc-p(C)VDZ and cc-pVDZ spectra in Fig. \ref{fig:basis}, we see the importance of the added core correlation functions. As expected, they cause a substantial shift in the probe absorption peaks, while they are not important for the pump absorption.
\begin{figure}
    \centering
    \includegraphics[width=3.375in]{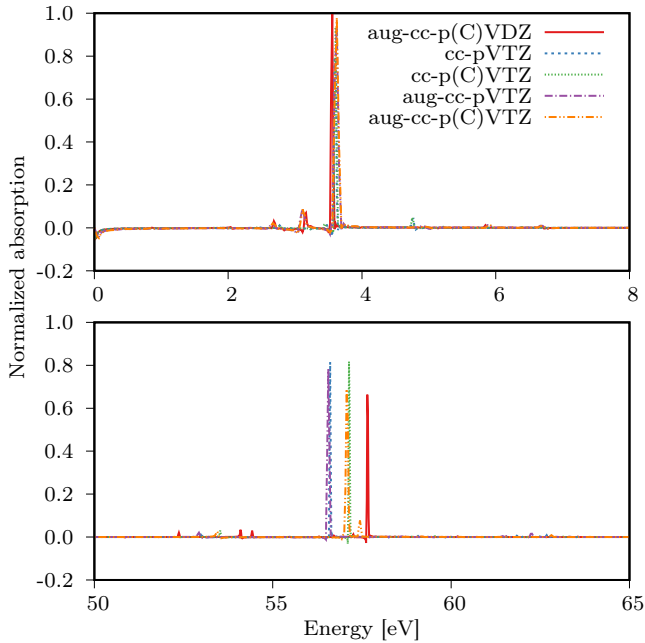}
    \caption{Normalized aug-cc-p(C)VDZ, cc-pVTZ, cc-p(C)VTZ, aug-cc-pVTZ and aug-cc-p(C)VTZ LiH pump-probe absorption, $S'(\omega)$, as a function of energy. Time-dependent dipole moments are calculated using TDCCSD, and integrated with RK4 with \SI{0.005}{\au} time steps.}
    \label{fig:basis_tz}
\end{figure}

We also performed calculations with cc-pVTZ, cc-p(C)VTZ, aug-cc-pVTZ and aug-cc-p(C)VTZ basis sets. Note that for the aug-cc-pVTZ and aug-cc-p(C)VTZ spectra, the time-dependent dipole moments are only calculated in the $[\SI{-2500}{\au},\SI{2500}{\au}]$ interval, in order to reduce computational time. Thus, these spectra have a lower resolution than the others. The normalized spectra are shown together with the normalized reference spectrum in Fig. \ref{fig:basis_tz}. Here we observe that triple zeta functions change the position of the peaks in the probe absorption region. This indicates that basis sets larger than aug-cc-p(C)VDZ should be used if precise peak positions are required, bringing about a substantial increase in the computational costs. The aug-cc-p(C)VDZ basis set is used as the reference for the other LiH calculations, as the larger basis sets are too computationally expensive for practical purposes.

\subsubsection{Integration}
\begin{figure}
    \centering
    \includegraphics[width=3.375in]{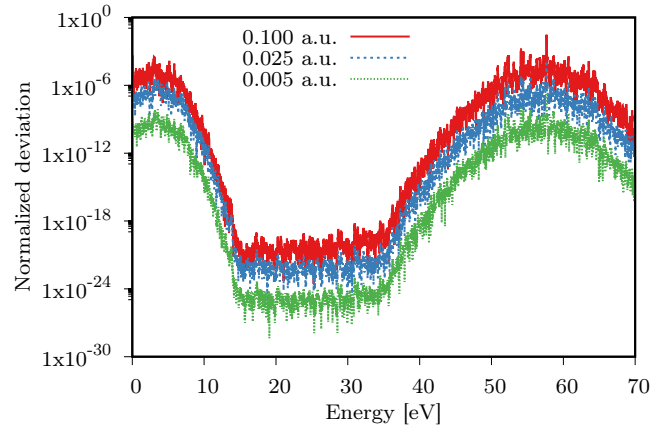}
    \caption{LiH pump-probe absorption. Normalized deviation of the \SI{0.125}{\au}, \SI{0.025}{\au} and \SI{0.005}{\au} time step spectra from the \SI{0.001}{\au} time step spectrum, $D'(\omega)$, as a function of energy. Time-dependent dipole moments are calculated using TDCCSD/aug-cc-p(C)VDZ, and integrated with RK4.}%
    \label{fig:timestep}
\end{figure}
%
%
\begin{figure}
    \centering
    \includegraphics[width=3.375in]{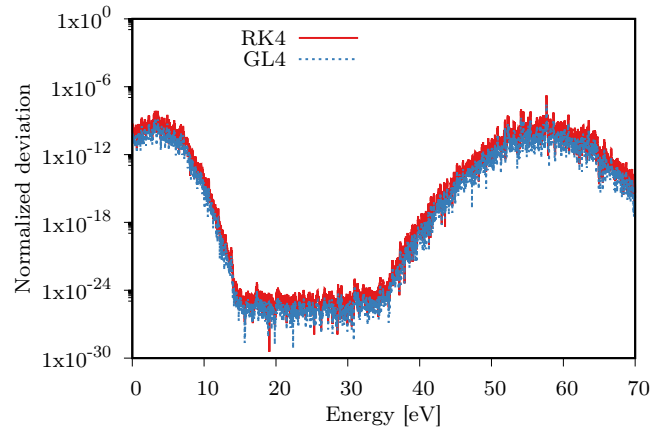}
    \caption{LiH pump-probe absorption. Normalized deviation of the RK4 and GL4 spectra from the GL6 spectrum, $D'(\omega)$, as a function of energy. Time-dependent dipole moments are calculated using TDCCSD/aug-cc-p(C)VDZ, and integrated with \SI{0.005}{\au} time steps.}
    \label{fig:integrator}
\end{figure}
We calculated normalized spectra for \SI{0.125}{\au}, \SI{0.025}{\au} and \SI{0.001}{\au} time steps. The deviations from the \SI{0.001}{\au} time step are calculated according to Eq. \eqref{eq:error}. The results are shown in Fig. \ref{fig:timestep}. The deviations decrease with the time step size, indicating that the spectra approach a time step limit.

We further calculated normalized spectra with GL4 and GL6. The deviations of the RK4 (reference) and GL4 spectra from the GL6 spectrum are shown in Fig. \ref{fig:integrator}. Although the TDCC equations have a Hamiltonian structure, the use of symplectic integrators does not seem to be necessary to calculate accurate spectra for this system, with the applied field strength. As the three integration methods give comparable results, we will use RK4 for the other calculations, as this generally requires fewer evaluations of the TDCC equations per time step.

\subsection{LiF transient absorption}
%
%
\begin{table}
    \centering
    \begin{ruledtabular}
    \begin{tabular}{lclc}
        State & $\omega$ [\si{\electronvolt}] & State & $\omega$ [\si{\electronvolt}]\T\BM \\
        \colrule
        ${\text{A}_{\text{v}}}^{1}\Pi$ & \num{6.44801} & ${\text{A}_{\text{c}}}^{1}\Sigma^{+}$ & \num{688.018}\T \\ 
        ${\text{B}_{\text{v}}}^{1}\Sigma^{+}$ & \num{6.89982} & ${\text{B}_{\text{c}}}^{1}\Pi$ & \num{689.462} \\
        ${\text{C}_{\text{v}}}^{1}\Delta$ & \num{8.10463} & ${\text{C}_{\text{c}}}^{1}\Sigma^{+}$ & \num{690.159} \\
        ${\text{D}_{\text{v}}}^{1}\Sigma^{-}$ & \num{8.14074} & ${\text{D}_{\text{c}}}^{1}\Sigma^{+}$ & \num{691.039} \\
        ${\text{E}_{\text{v}}}^{1}\Sigma^{+}$ & \num{8.51116} & ${\text{E}_{\text{c}}}^{1}\Pi$ & \num{691.435} \\
        ${\text{F}_{\text{v}}}^{1}\Pi$ & \num{8.58943} & ${\text{F}_{\text{c}}}^{1}\Sigma^{+}$ & \num{691.625} \\
        ${\text{G}_{\text{v}}}^{1}\Pi$ & \num{8.62589} & ${\text{G}_{\text{c}}}^{1}\Pi$ & \num{692.917} \\
        ${\text{H}_{\text{v}}}^{1}\Sigma^{+}$ & \num{9.10655} & ${\text{H}_{\text{c}}}^{1}\Sigma^{+}$ & \num{693.154}
    \end{tabular}
    \end{ruledtabular}
    \caption{Molecular term symbols and ground state excitation energies of some excited states of LiF, calculated with the EOM-CCSD method. Valence-excited states are denoted by a subscript v. Core-excited states, calculated within the CVS approximation, are denoted by a subscript c.}
    \label{tab:states}
\end{table}
\begin{figure}
    \centering
    \includegraphics[width=3.375in]{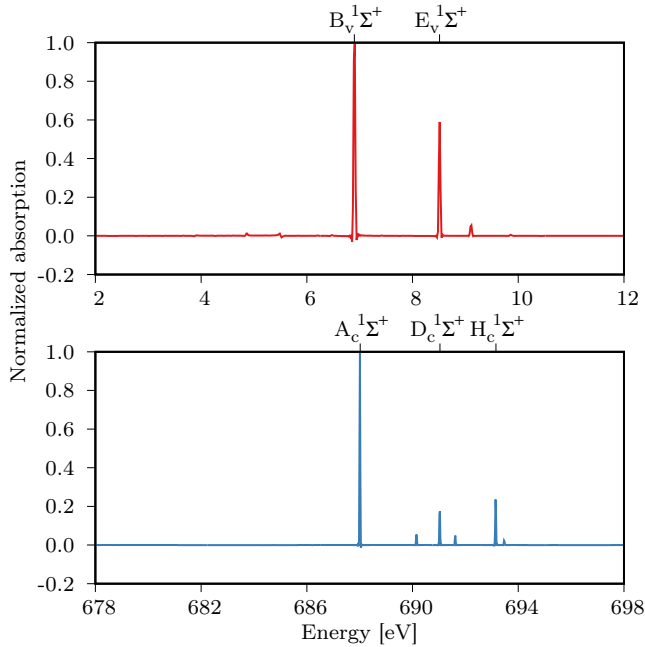}
    \caption{Normalized LiF pump and probe absorption, $S_{\text{pump}}'(\omega)$ (top) and $S_{\text{probe}}'(\omega)$ (bottom), as a function of energy. The most dominant peaks are identified with ground state transitions to EOM-CCSD valence- and core-excited states. Time-dependent dipole moments are calculated with TDCCSD/aug-cc-p(C)VDZ, integrated with RK4 with \SI{0.005}{\au} time steps.}
    \label{fig:pumpandprobe}
\end{figure}
\begin{figure*}
    \centering
    \includegraphics[width=6.75in]{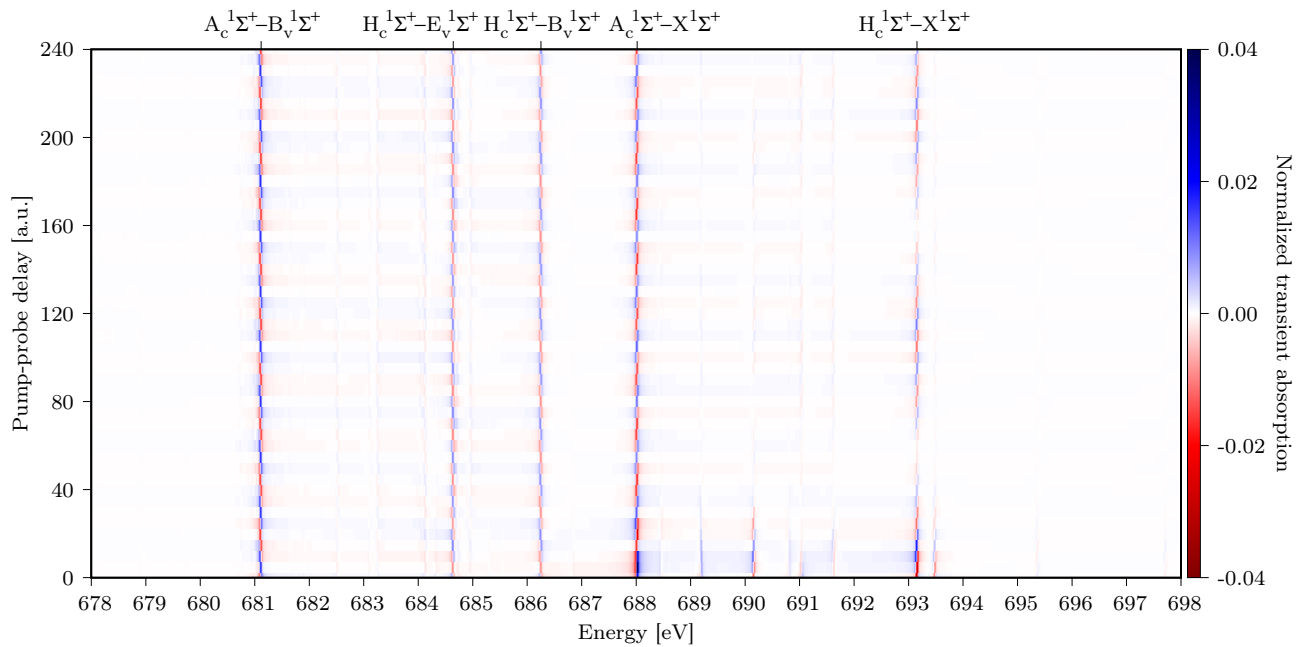}
    \caption{Normalized LiF transient absorption $\Delta S'(\omega,\tau)$, as a function of energy and pump-probe delay. The five peaks oscillating with the largest amplitude are identified with EOM-CCSD transitions. Time-dependent dipole moments are calculated using TDCCSD/aug-cc-p(C)VDZ basis set, integrated with RK4 with \SI{0.005}{\au} time steps.}
    \label{fig:transient_2d}
\end{figure*}
\begin{figure}
    \centering
    \includegraphics[width=3.375in]{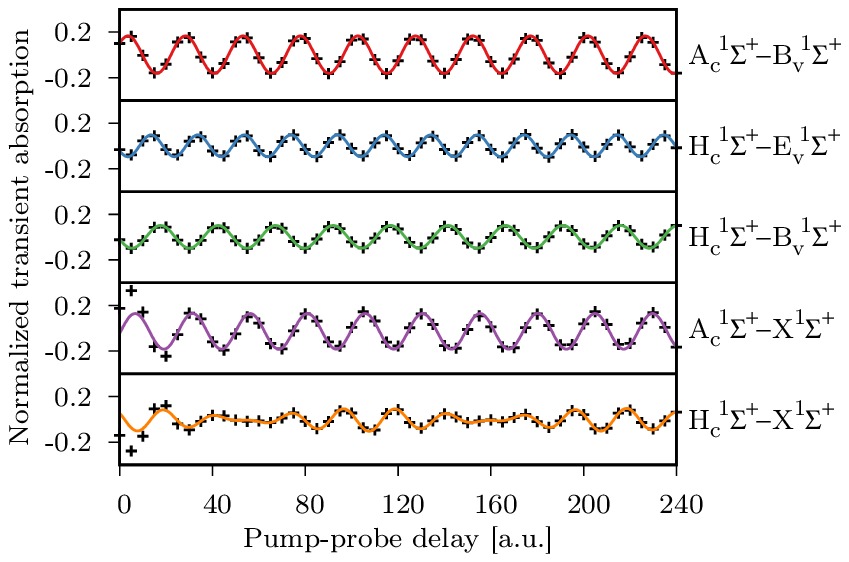}
    \caption{Normalized LiF transient absorption $\Delta S'(\omega,\tau)$ (black crosses) as a function of pump-probe delay, given at the discrete Fourier transform energies closest to the energies of the transitions shown to the right. The colored functions in the four topmost panels are found from least-squares fitting $A\sin(\omega_{A}t+\phi_{A})+C$, with fixed values of $\omega_{A}$, to the absorption, in the domain $[\SI{40}{\au},\SI{240}{\au}]$. The values of $\omega_{A}$ are $\SI{6.89982}{\electronvolt}$ (red) $\SI{8.51116}{\electronvolt}$ (blue), $\SI{6.89982}{\electronvolt}$ (green) and $\SI{6.89982}{\electronvolt}$ (purple). The orange function in the bottom panel is found from least-squares fitting $A\sin(\omega_{A} t+\phi_{A})+B\sin(\omega_{B} t+\phi_{B})+C$, with $\omega_{A}=\SI{6.89982}{\electronvolt}$ and $\omega_{B}=\SI{8.51116}{\electronvolt}$, to the absorption, in the domain $[\SI{40}{\au},\SI{240}{\au}]$.}
    \label{fig:modulation}
\end{figure}
In this section, variations in molecular absorption caused by ultrafast charge migration are modelled in the described pump-probe framework. We consider the lithium fluoride (LiF) molecule, where the fluorine atom is placed at the origin and the lithium atom at \SI{-1.56386413}{\angstrom} along the $z$-axis. This corresponds to the experimentally measured equilibrium bond length of LiF \cite{cccbdb}. In order to classify some of the transitions involved in the molecular absorption, the first eight valence-excited and the first eight core-excited states are calculated using EOM-CCSD/aug-cc-p(C)VDZ. The core excitations are obtained within the CVS approximation. The molecular term symbols and excitation energies are given in Table \ref{tab:states}.

In the TDCC calculations, all probe pulses are $z$-polarized, and have a central angular frequency corresponding to the first LiF valence-excitation energy (see Table \ref{tab:states}). Central times are chosen to be \SI{0}{\au}, to minimize the effect of the windowing on the probe absorption. The pump pulses are also $z$-polarized, and have a central angular frequency corresponding to the first LiF core-excitation energy (see Table \ref{tab:states}). The pump pulses have different central times with respect to the probe pulses, corresponding to probe delays from \SI{0}{\au} to \SI{240}{\au}, in increments of \SI{5}{\au} Other parameters of the pump and probe pulses are given in Table \ref{tab:pulse_parameters}. As for the LiH calculations, the electric fields of each pulse are temporally truncated at eight RMS widths $\sigma$ from the central time, and thus nonzero only inside this interval (see Eq. \eqref{eq:envelope}).

The parameters used for the LiH reference calculation offered a compromise between computational cost and accuracy. For pragmatic reasons, we also use the parameters for all LiF calculations. The calculations in this section are thus done using TDCCSD/aug-cc-p(C)VDZ, and integrated with RK4 with \SI{0.005}{\au} time steps. The time-dependent dipole moments and electric fields are calculated every \SI{0.1}{\au} in the $[\SI{-5000}{\au}, \SI{5000}{\au}]$ interval, where the external field interactions are only calculated after the onset of the temporally truncated pump pulses.

In order to assess the relative occupation of the states in the pump-induced superposition (see Eq. \eqref{eq:transientenergy}), the normalized absorption of the pump pulse, centered at \SI{0}{\au}, is calculated using
\begin{equation}
    S_{\text{pump}}'(\omega) = \mathcal{N}_{\text{pump}}S_{\text{pump}}(\omega),
\end{equation}
where
\begin{equation}
    \mathcal{N}_{\text{pump}} = \frac{1}{\max_{\omega}\abs{S_{\text{pump}}(\omega)}}.
\end{equation}
An analogous procedure is used to obtain the normalized probe spectrum $S_{\text{probe}}'(\omega)$.

The normalized absorption of the pump pulse, and of the probe, pulse are plotted in Fig. \ref{fig:pumpandprobe}, where the most dominant absorption peaks are identified using the calculated EOM-CCSD states (see Table \ref{tab:states}). The small pump absorption peaks that lie below the ground state valence-excitation energy gap are presumably caused by two-photon absorption. The positions of the other visible peaks in the two spectra fit well with single-photon EOM-CCSD transitions allowed by symmetry.

The pump-probe absorption $S(\omega,\tau)$ is calculated as a function of the energy, $\omega$, and the delay of the probe pulse with respect to the pump pulse, $\tau$. In order to directly assess the change in absorption caused by the interaction with the pump pulse, the normalized transient absorption
\begin{equation}
    \begin{split}
        \Delta S'(\omega, \tau) &= \mathcal{N}_{\text{probe}} \Delta S(\omega, \tau) \\ &= \mathcal{N}_{\text{probe}}\Big(S(\omega, \tau) - S_{\text{probe}}(\omega)\Big),
    \end{split}
\end{equation}
is calculated for all delays, where $\mathcal{N}_{\text{probe}}$ is the normalization factor for the probe spectrum. The normalized transient absorption in the probe absorption region is shown in Fig. \ref{fig:transient_2d}. The spectrum features several constant energy peaks that oscillate with the pump-probe delay. The five peaks that oscillate the most with respect to the pump-probe delay are identified using the states in Table \ref{tab:states}. Note that for shorter pump-probe delays, the oscillations of some of the peaks are rapidly damped as a function of increasing delays. This effect can be attributed to the decreasing overlap between the pump and probe pulses. For longer pump-probe delays, where the overlap of the pulses is negligible, the oscillations are undamped.

We note that the excitation by the pump pulse enables new transitions in the probe absorption region. An illustrative example is the oscillating peak at around \SI{681.1}{\electronvolt} in Fig. \ref{fig:transient_2d}. The energy corresponding to this peak is lower than the lowest ground state core-excitation energy of \SI{688.018}{\electronvolt}. This peak is identified as the ${\text{A}_{\text{c}}}^{1}\Sigma^{+}$--${\text{B}_{\text{v}}}^{1}\Sigma^{+}$ transition. Its occurrence indicates that the pump has generated an electronic hole in a previously occupied region of the molecule, allowing a lower energy core excitation to take place.

In Fig. \ref{fig:modulation}, the normalized transient absorption of the five peaks identified in Fig. \ref{fig:transient_2d} are plotted at the nearest discrete Fourier transform energies. Two of these peaks describe transitions involving the ${\text{A}_{\text{c}}}^{1}\Sigma^{+}$ state. Beyond the pump-probe overlap region, the oscillations of these peaks correlate with the quantum interference of the two probed states, as expected for the ultrafast high-energy probing of two states in a coherent superposition \cite{Goulielmakis2010}. This since both oscillations can be fitted with sinusoids with the frequency corresponding to the ${\text{B}_{\text{v}}}^{1}\Sigma^{+}$ and ${\text{X}}^{1}\Sigma^{+}$ energy difference.

Three peaks in Fig. \ref{fig:modulation} correspond to transitions involving the ${\text{H}_{\text{c}}}^{1}\Sigma^{+}$ state. The oscillation of the ${\text{H}_{\text{c}}}^{1}\Sigma^{+}$--${\text{B}_{\text{v}}}^{1}\Sigma^{+}$ peak correlates well with the quantum interference of the ${\text{B}_{\text{v}}}^{1}\Sigma^{+}$ and ${\text{X}}^{1}\Sigma^{+}$ states, as the oscillations are well fitted with a sinusoids with frequency corresponding to the energy difference of these two states. Similarly, the oscillation of the ${\text{H}_{\text{c}}}^{1}\Sigma^{+}$--${\text{E}_{\text{v}}}^{1}\Sigma^{+}$ peak correlates with the quantum interference of the ${\text{E}_{\text{v}}}^{1}\Sigma^{+}$ and ${\text{X}}^{1}\Sigma^{+}$ states. Note that the oscillations of the two peaks are slightly phase shifted with respect to each other, an effect that may be caused by the difference in spectral phase of the two corresponding frequencies in the probe pulse.

The linear combination of two sinusoids is needed to give a good fit with the oscillation of the ${\text{H}_{\text{c}}}^{1}\Sigma^{+}$--${\text{X}}^{1}\Sigma^{+}$ peak: one corresponding to the ${\text{B}_{\text{v}}}^{1}\Sigma^{+}$ and ${\text{X}}^{1}\Sigma^{+}$ energy difference, and the other corresponding to the ${\text{E}_{\text{v}}}^{1}\Sigma^{+}$ and ${\text{X}}^{1}\Sigma^{+}$ energy difference. Hence, the ground state ${\text{X}}^{1}\Sigma^{+}$ seems to have a similar probability of interfering with the ${\text{B}_{\text{v}}}^{1}\Sigma^{+}$ and ${\text{E}_{\text{v}}}^{1}\Sigma^{+}$ states. This is reasonable, considering that most of the population will be left in the ground state after the interaction with the pump pulse.
\section{Conclusion} \label{sec:conclusion}
In this work, a time-dependent coupled cluster model of ultrafast pump-probe absorption spectroscopy has been presented. First, we investigated the convergence of LiH pump-probe absorption spectra with respect to different calculation parameters. The deviations related to the integration parameters (integration method and time step size) were small in comparison to other parameter-dependent deviations. As the computational costs scaled linearly with the time step size, we chose a time step size that gave a small deviation, \SI{0.005}{\au} The use of symplectic integrators did not seem to be necessary, hence RK4 was used. Changes in the basis set had a big impact on the results. As the computational cost scales steeply with respect the basis set, TDCCSD/aug-cc-p(C)VDZ was chosen as a compromise between accuracy and computational cost.

After using the time-dependent coupled cluster model to assess the convergence of LiH spectra, we used the model to calculate the ultrafast transient absorption in LiF, using the same parameters. The transient absorption displayed peaks that oscillate with respect to pump-probe delay, and the oscillation frequencies were correlated with the quantum interference of different states in the pump-induced superposition.
%
\begin{acknowledgments}
%
We acknowledge the financial support from The Research Council of Norway through FRINATEK Project Nos. 263110 and 275506, and computing resources through UNINETT Sigma2---the National Infrastructure for High Performance Computing and Data Storage in Norway (Project No. NN2962k) and through the SMART@SNS Laboratory.
\end{acknowledgments}
%
\appendix
%
\section{Derivation of generalized Ehrenfest theorem in truncated TDCC}
%
For ease of notation, the time dependence is not written explicitly in this section. The derivation of Eq. \eqref{eq:ehrenfest} in truncated TDCC is given here. It makes use of the identity resolution
\begin{equation}
    \label{eq:resolution}
    \mathbb{1} = \ket{\HF}\!\bra{\HF} + \sum_{\mu>0}\ket{\mu}\!\bra{\mu},
\end{equation}
where the summation is over all the excited determinants. Sums that are restricted to the excited determinants in the projection space, will be denoted by the upper summation limit $n$.

Consider a generic operator $A$ with no parametric time dependence, and two independent solutions to the projected time-dependent Schr\"{o}dinger equation, $\ket{\CC}$ and $\bra{\Lambda'}$. The time derivative of the matrix element $\mel{\Lambda '}{A}{\CC}$ is
\begin{equation}
    \label{eq:ddtexp}
    \begin{split}
        \dv{t}\mel{\Lambda'}{A}{\CC} &= \bigg(\dv{t}\bra{\Lambda'}\!\bigg)A\!\ket{\CC} + \mel{\Lambda'}{\pdv{A}{t}}{\CC} \\
        &\phantom{{}={}}
        + \bra{\Lambda'}\!A\bigg(\dv{t}\ket{\CC}\!\bigg).
    \end{split}
\end{equation}
Equations \eqref{eq:ddtmultipliers}, \eqref{eq:ddtamplitudes} and \eqref{eq:ddtphase} can be used to rewrite the term containing the time derivative of the $\bra{\Lambda'}$ state,
\begin{equation}
    \label{eq:ddtlambdap1}
    \begin{split}
       &\bigg(\dv{t}\bra{\Lambda'}\!\bigg)A\ket{\CC} \\
       &= \sum_{\mu>0}^{n}\dv{\tbar_{\mu}'}{t}\mel{\mu}{e^{-T'}e^{-i\epsilon'}A}{\CC} \\
       &\phantom{{}={}}-\sum_{\mu>0}^{n} \mel{\Lambda'}{\tau_{\mu}A}{\CC}\dv{t_{\mu}'}{t}- i \mel{\Lambda'}{A}{\CC}\dv{\epsilon'}{t} \\
        &=
        \sum_{\mu>0}^{n}i\mel{\Lambda'}{He^{T'}}{\mu}\!\mel{\mu}{e^{-T'}A}{\CC} \\ &\phantom{{}={}}
        - \sum_{\mu>0}^{n}i\mel{\Lambda'}{e^{T'}\tau_{\mu}\Hbar'}{\HF}\!\mel{\mu}{e^{-T'}A}{\CC} \\
        &\phantom{{}={}} +\sum_{\mu>0}^{n}i\mel{\Lambda'}{\tau_{\mu}A}{\CC}\!\mel{\mu}{\Hbar'}{\HF} \\&\phantom{{}={}}
        +i \mel{\Lambda'}{A}{\CC}\!\mel{\HF}{\Hbar'}{\HF}.
    \end{split}
\end{equation}
The right hand side of equation \eqref{eq:resolution} is inserted between $\tau_{\mu}$ and $\Hbar'$ in the second term, giving
\begin{equation}
    \label{eq:ddtlambdap2}
    \begin{split}
        &\bigg(\dv{t}\bra{\Lambda'}\!\bigg)A\ket{\CC} \\
        &=\sum_{\mu>0}^{n}i\mel{\Lambda'}{He^{T'}}{\mu}\!\mel{\mu}{e^{-T'}A}{\CC} \\
        &\phantom{{}={}} - \sum_{\mu>0}^{n}i\mel{\Lambda'}{e^{T'}}{\mu}\!\mel{\HF}{\Hbar'}{\HF}\!\mel{\mu}{e^{-T'}A}{\CC} \\
        &\phantom{{}={}} - \!\!\sum_{\mu>0}^{n}\sum_{\nu>0}\!i\!\mel{\Lambda'}{e^{T'}\tau_{\mu}}{\nu}\!\!\mel{\nu}{\Hbar'}{\HF}\!\!\mel{\mu}{e^{-T'}A}{\CC} \\
        &\phantom{{}={}} +\sum_{\mu>0}^{n}i\mel{\Lambda'}{\tau_{\mu}A}{\CC}\!\mel{\mu}{\Hbar'}{\HF} \\
        &\phantom{{}={}}
        +i \mel{\Lambda'}{A}{\CC}\!\mel{\HF}{\Hbar'}{\HF} \\
        &=\sum_{\mu>0}^{n}i\mel{\Lambda'}{He^{T'}}{\mu}\!\mel{\mu}{e^{-T'}A}{\CC} \\
        &\phantom{{}={}} - \sum_{\mu>0}i\mel{\Lambda'}{e^{T'}}{\mu}\!\mel{\mu}{e^{-T'}A}{\CC}\!\mel{\HF}{\Hbar'}{\HF} \\
        &\phantom{{}={}} - \!\!\sum_{\nu>0}^{n}\sum_{\mu>0}\!i\!\mel{\Lambda'}{e^{T'}\tau_{\nu}}{\mu}\!\!\mel{\mu}{e^{-T'}A}{\CC}\!\!\mel{\nu}{\Hbar'}{\HF} \\
        &\phantom{{}={}} +\sum_{\mu>0}^{n}i\mel{\Lambda'}{\tau_{\mu}A}{\CC}\!\mel{\mu}{\Hbar'}{\HF} \\
        &\phantom{{}={}}
        +i \mel{\Lambda'}{A}{\CC}\!\mel{\HF}{\Hbar'}{\HF} \\
    \end{split}
\end{equation}
The factors $\sum_{\mu>0}\ket{\mu}\!\bra{\mu}$ in the second and third terms are replaced by using Eq. \eqref{eq:resolution}, with $\ket{\HF}\!\bra{\HF}$ subtracted from both sides of the equation, giving
\begin{equation}
    \label{eq:ddtlambdap3}
    \begin{split}
        &\bigg(\dv{t}\bra{\Lambda'}\!\bigg)A\ket{\CC} \\
        &=\sum_{\mu>0}^{n}i\mel{\Lambda'}{He^{T'}}{\mu}\!\mel{\mu}{e^{-T'}A}{\CC} \\
        &\phantom{{}={}} + i\mel{\Lambda'}{e^{T'}}{\HF}\!\mel{\HF}{e^{-T'}A}{\CC}\!\mel{\HF}{\Hbar'}{\HF} \\
        &\phantom{{}={}} + \sum_{\nu>0}^{n}i\mel{\Lambda'}{e^{T'}}{\nu}\!\mel{\HF}{e^{-T'}A}{\CC}\!\mel{\nu}{\Hbar'}{\HF} \\
        &=\sum_{\mu>0}^{n}i\mel{\Lambda'}{He^{T'}}{\mu}\!\mel{\mu}{e^{-T'}A}{\CC} \\
        &\phantom{{}={}} + i\mel{\Lambda'}{e^{T'}}{\HF}\!\mel{\HF}{\Hbar'}{\HF}\!\mel{\HF}{e^{-T'}A}{\CC} \\
        &\phantom{{}={}} + \sum_{\nu>0}i\mel{\Lambda'}{e^{T'}}{\nu}\!\mel{\nu}{\Hbar'}{\HF}\!\mel{\HF}{e^{-T'}A}{\CC}\\
        &=
        i\mel{\Lambda'}{He^{T'}P_{n}e^{-T'}A}{\CC},
    \end{split}
\end{equation}
where definition of $P_{n}$ is given in Eq. \eqref{eq:projection}.
Equations \eqref{eq:ddtamplitudes} and \eqref{eq:ddtphase} can also be used to rewrite the term containing the time derivative of the $\ket{\CC}$ state,
\begin{equation}
    \label{eq:ddtcc}
    \begin{split}
        &\bra{\Lambda'}A\bigg(\dv{t}\ket{\CC}\!\bigg) \\
        &= \sum_{\mu>0}^{n}\mel{\Lambda'}{A\tau_{\mu}}{\CC}\dv{t_{\mu}}{t} + i \mel{\Lambda'}{A}{\CC}\dv{\epsilon}{t} \\
        &= -i\mel{\Lambda'}{A e^{T}P_{n}e^{-T} H}{\CC}.
    \end{split}
\end{equation}
Eqs. \eqref{eq:ddtlambdap3} and \eqref{eq:ddtcc} are inserted into Eq. \eqref{eq:ddtexp}, giving the desired result
\begin{equation}
    \begin{split}
        \dv{t}\mel{\Lambda'}{A}{\CC} &= i\mel{\Lambda'}{He^{T'}P_{n}e^{-T'}A}{\CC} \\
        &\phantom{{}={}} -i\mel{\Lambda'}{Ae^{T}P_{n}e^{-T}H}{\CC} \\
        &\phantom{{}={}}+ \mel{\Lambda'}{\pdv{A}{t}}{\CC}.
    \end{split}
\end{equation}
\section{Runge-Kutta methods}
%
The commonly used one-step integration methods known as Runge-Kutta methods are introduced below in the notation of \cite{iserles_2008}.

Given the following Cauchy problem
\begin{equation}
	\dv{\vb{y}(t)}{t} = \vb{f}(t,\vb{y}(t)) \qc t \ge t_{0} \qc \vb{y}(t_{0})=\vb{y}_{0},
\end{equation}
we can find a numerical approximation of the solution $\vb{y}(t)$ by the use of a $\nu$-stage Runge-Kutta method, which can be written in the form
\begin{equation}
    \vb{y}_{n+1} = \vb{y}_{n} + h \sum_{j=1}^{\nu} b_{j} \vb{f}\Big(t_{n} + c_{j}h, \vb*{\xi}_{j}\Big),
\end{equation}
where
\begin{equation}
    \vb*{\xi}_{j} = \vb{y}_{n} + h\sum_{i=1}^{\nu} a_{ji} \vb{f} \Big(t_{n} + c_{i} h , \vb*{\xi}_{i}\Big) \qc j = 1, \ldots, \nu.
\end{equation}
Here, $a_{ji}$, $b_{j}$ and $c_{j}$ are method specific coefficients, where $a_{ji}$ and $c_{j}$ need to satisfy the condition
\begin{equation}
    \sum_{j=1}^{\nu}a_{ji} = c_{j}\qc j = 1, \ldots, \nu
\end{equation}
to obtain non-trivial orders of integration. In explicit Runge-Kutta (ERK) methods, the matrix $A = (a_{ji})_{j,i=1,\ldots,\nu}$ is strictly lower triangular. In these methods, $\vb*{\xi}_{j}$ are explicitly given as a function of $\vb*{\xi}_{j-1}, \ldots, \vb*{\xi}_{1}$.

In the cases where the matrix $A$ is not strictly lower triangular, $\vb*{\xi}_{j}$ may also depend on $\vb*{\xi}_{j}, \ldots, \vb*{\xi}_{\nu}$, which in practice means that a system of equations have to be solved at each time step. These methods are known as implicit Runge-Kutta (IRK) methods, and in many cases offer greater stability than their explicit counterparts. Since IRK methods involve the solution of a set of equations at each time step, it is hard to give an \textit{a priori} estimate of the number of function evaluations needed at each time step. This number is usually higher than for ERK methods, leading in general to higher computational costs.

\bibliographystyle{apsrev4-2}
\bibliography{bibliography}
\end{document}